\documentclass[a4paper,preprint,unsortedaddress]{revtex4-1}
\usepackage{amsmath,amssymb,amsfonts,latexsym}
\usepackage[dvips]{graphicx}
\usepackage{color}
\usepackage{indentfirst}

\newcommand{\redc}[1]{{\color{black} #1}}            
            
\newcommand{\vect}[1]{\textbf{\textit{#1}}}

\newcommand{\AT}{{\textrm{{AT}}}}

\newcommand{\CG}{{\textrm{CG}}}
\newcommand{\HY}{{\Delta}}

\newcommand{\thf}{{\textrm{th}}}

\newcommand{\thermo}{{\textrm{Q}}}

\begin{document}
\title{Molecular Dynamics in a Grand Ensemble: Bergmann-Lebowitz model and Adaptive Resolution Simulation}
\author{Animesh Agarwal}
\affiliation{Institute for Mathematics, Arnimallee 6, D-14195, Freie Universit\"{a}t, Berlin, Germany}
\author{Jinglong Zhu}
\affiliation{School of Mathematical Sciences, Peking University, Beijing, 100871
P. R. China}
\author{Carsten Hartmann}
\affiliation{Institute for Mathematics, Arnimallee 6, D-14195, Freie Universit\"{a}t, Berlin, Germany}
\author{Han Wang}
\affiliation{CAEP Software Center for High Performance Numerical Simulation, Beijing, P. R. China}
\author{Luigi Delle Site}
\email{luigi.dellesite@fu-berlin.de}
\affiliation{Institute for Mathematics, Arnimallee 6, D-14195, Freie Universit\"{a}t, Berlin, Germany}
\begin{abstract}
\redc{This article deals with the molecular dynamics simulation of open systems that can exchange energy and matter with a reservoir; the physics of the reservoir and its interactions with the system are described by the model introduced by Bergmann and Lebowitz. \redc{Despite its conceptual appeal, the model did not gain popularity in the field of molecular simulation and, as a consequence, did not play a role in the development of open system molecular simulation techniques}, even though it can provide the conceptual legitimation of simulation techniques that mimic open systems. We shall demonstrate that the model 
can serve as a tool to devise both numerical procedures and conceptual definitions of physical quantities that cannot be defined in a straightforward way by  systems with a fixed number of molecules. \redc{In particular, we discuss the utility of the Bergmann-Lebowitz (BL) model for the calculation of equilibrium time correlation functions within the Grand Canonical Adaptive Resolution method (GC-AdResS) and report numerical results for the case of liquid water}.}

\end{abstract}

\maketitle
\section{Introduction}
\redc{The physics of open systems is considered of primary importance in the understanding of natural phenomena and in the development of modern technology \cite{blatt}}. Systems in real life as well as \redc{in} experimental \redc{set-up}s are open systems, that is, systems which exchange energy and particles with their environment; the \redc{process of exchange of particles is at the basis of their interesting properties (see e.g.\cite{cell}). From a theoretical point of view the conceptual development of the classical and quantum statistical mechanics of open systems is challenging; in fact theorems of statistical mechanics and dynamics derived for systems with a fixed number of particles are no longer valid in their standard formulation and must be} \redc{revised accordingly, e.g., if the deterministic evolution is substituted by the stochastic evolution which controls the process of exchange of particles \cite{leb1,leb2,leb3,jmp,peier}.  We will argue that extensive theoretical work with effective, elegant and (from a practical point of view) useful concepts have been developed a long time ago}  \redc{(much in advance with respect to the advent of computer simulations) but \redc{have} remained unnoticed by the majority of the molecular simulation community}. As a matter of fact, until recently,  open systems with varying number of particles have been simulated using algorithms which did not succeed as expected. The lack of success was most probably due to reduced efficiency compared to techniques based on fixed particle numbers (see, e.g.\cite{pettit1}). However, recently \redc{algorithms of multiscale character, which aim at bridging different scales within one \redc{unified} framework, have gained large popularity, which in turn has led to the construction of efficient techniques where systems exchange energy \redc{or particles} with an external environment.} \redc{For techniques using} molecular resolution that can adaptively change in space (adaptive resolution simulation, see e.g. \cite{entropy} and \cite{annurev} and references therein).

Adaptive resolution simulation techniques allow to focus on a specific region in space, treated at a desired (high) resolution, while the rest of the system is treated at a lower resolution. In the resolved region, some interesting process takes place while the rest of the system stays in thermodynamic equilibrium with the subsystem of interest (or, beyond equilibrium, exchanges energy and particles according to well defined statistical physical laws). In contrast to the first generation of algorithms with varying number of particles, such algorithms are technically highly efficient and flexible. \redc{This flexibility makes them feasible for the calculation of various statistical properties, such as time correlation functions, some of which require a theoretical re-definition (compared to the fixed particle number simulations). The necessity of a formal re-definition of equilibrium time correlation functions in modern open systems MD simulations calls for revisiting the theoretical concepts developed about 5-6 decades ago for the statistical mechanics of open systems in the context of state-of-the-art computer algorithms.} 

In this paper, following the terminology developed in \cite{leb1,leb2,leb3}, we will refer to open systems which exchange energy and matter with the environment as Grand Ensemble systems; the Grand Canonical Ensemble is one particular realization of a Grand Ensemble, as discussed in \cite{leb1,leb2,leb3}. 
The aim of this paper is: (a) a discussion of theoretical concepts of open systems present in literature; (b) a brief overview about the development/application of algorithms with varying number of particles in Molecular Dynamics; (c) the inclusion/adaptation of formal results about open systems into the framework of MD techniques; (d) \redc{to provide} examples of merging theory and algorithms by reporting numerical results for one specific open system MD technique. We will treat the specific case of Grand Canonical-like Adaptive Resolution Simulation method (GC-AdResS) and discuss its conceptual consistency with the theory present in literature together with its technical advantages/limitations. 
\redc{The hope is that this may stimulate further research along this direction and \redc{add} to the theoretical foundation of MD simulations in a Grand Ensemble; the need of approaching more complex systems characterized by the realistic process of exchange of energy and matter with the environment, prohibitive in the past, is becoming a guiding principle in the development and the application of Molecular Simulation techniques \cite{parrix}.\redc{set-up}} 

The paper is organized as follows: in the first section we will give a general overview of the theoretical concepts developed about the statistical mechanics of open systems. Next we will focus on what we will call the Bergmann-Lebowitz approach, \redc{a flexible and conceptually robust model that is of utmost relevance for many state-of-the-art MD algorithms}. In the second section we will briefly discuss the general features of techniques of MD with varying number of particles and introduce the idea of MD with molecular adaptive resolution simulation. In the third section we will introduce one of the techniques of adaptive resolution simulation (GC-AdResS) and report results where the BL theory of the first section is employed to give conceptual justification \redc{of} the simulations and to the corresponding calculation technique. Finally conclusions and future perspectives will be given. Finally, it must be noticed that the technical \redc{set-up} and the numerical results reported in this work are original in the development of the GC-AdResS methodology. In fact the results show that, with the technical \redc{set-up} developed in this work, the method is reliable not only for the calculation of static properties, on which past research was focused, but also for the calculation of dynamical properties, thus allowing the study of a much larger class of phenomena.

\section{Basic concepts of a Grand Ensemble and Extended Liouville Equation}
\redc{ When one uses the keyword {\it ``statistical mechanics of open systems''} \redc{in an automatic literature search}}, one finds a considerable amount of rich material (see, e.g., Refs.\cite{leb1,leb2,leb3,jmp,peier,seke}). However the vast majority of this material focuses mostly on the idea of coupling a system to a reservoir of energy, or to non-equilibrium scenarios, such as the transport of matter from an external source. \redc{Exchange of matter techniques are usually limited to a simple extensions of the concept of heat exchange and heat flow \cite{jmp,leb2,leb3}. As a matter of fact, the exchange of heat has been historically most relevant in MD simulations, as the coupling a system to an external reservoir of energy makes simulations numerically stable and physically targeted to the desired thermodynamic state,} without requiring large systems as those necessary to NVE simulations \cite{tuck}. The circumstances outlined above, together with the lack of success of Grand Canonical-like MD methods (see the later discussion), were the reasons why the theoretical concepts of Grand Ensemble, developed, e.g.,in Refs.\cite{leb2,leb3,jmp} did not become popular in MD simulations and thus \redc{were not implemented in } practical tools of calculations. \redc{However,  as underlined in the introduction, the re-discovery and further development of such work became a timely necessity}. In this section we will trace the idea behind the theoretical treatment of open systems in equilibrium and we will restrict the discussion to those approaches where the \redc{coupling between system and reservoir is not required in an explicit form; such approaches represent the most general model open system}.  Moreover, we will restricted the treatment to classical systems because our main interest lies in the field of classical MD. In particular we will define a generalized Liouville equation and associated operator (the Bergmann-Lebowitz Liouville equation/operator). 
\redc{Instead, the class of approaches which explicitly require a coupling term in the Hamiltonian is usually limited to transport processes (out of equilibrium), whose  external source can be formalized in specific cases only (e.g.~\cite{tech}) and which in general do not admit a Grand Ensemble.}
The essential idea behind \redc{approaches which do not require an explicit coupling term, is that a small system is in contact a large reservoir (or more than one, but for simplicity let us consider only one). The aim is to extract (thermo)dynamical laws governing the small system, from the microscopic equation of the global system (comprising the reservoir)}.  The Liouville equation of the global system is the ideal starting point, however, the variables considered in the reservoir are macroscopic variables \redc{that can be considered averages over microscopic states; in the optimal case these variables do not explicitly  enter in the description of the evolution of the small system.}
\redc{The general hypothesis at the basis of such models is that the reservoir exerts its influence on the small system only via intensive properties (see e.g.\cite{jmp,leb2})}. The key idea is that, even if the extensive variables of the reservoir change, its intensive variables are constants of motion. As a consequence the dynamical evolution of the small system does not contain any time-dependent function of the reservoir and the small system is then governed by a self-\redc{consistent} dynamical evolution. In a pioneering work, Emch and Sewell \cite{jmp} proposed a method based on the basic principles reported before. They treat quantum systems, and the generalized Liouville equation is a master equation governing the evolution of the statistical operator. However, they need an abstract projector operator which coarse-grains the microscopic variables of the reservoir into macroscopic variables, \redc{that, in turn, influence the small subsystem. For MD simulations, although the premises of the method and its formalism are certainly appealing, this idea is not practical, in fact the explicit specification and formalization of a general coarse-graining operator is not straightforward. However, a similar, but more appealing idea for its formal simplicity and from the viewpoint of practical implementation, has been put forward by Bergman and Lebowitz \cite{leb2} as will be outlined below; see also Ref.\cite{leb3}}.

\subsection{Bergmann-Lebowitz Liouville equation}
In the seminal paper of Bergmann and Lebowitz \cite{leb2} (and subsequently in the paper of Lebowitz and Shimony \cite{leb3}), \redc{the authors derive a general model of a many-particle system that is interacting with different reservoirs.} Here, for simplicity and for closer analogy to a standard Grand Canonical MD simulations, we will treat only the case of a single reservoir. \redc{The key ingredient of the model is an impulsive, Markovian interaction between the reservoir and the system. The  effect of the reservoir on the system can be completely described if one specifies the stationary distribution of the reservoir before the reservoir-system interaction (thus the knowledge of the reservoir state as a function of time is not required).}
In their model, each interaction between the system and the reservoir produces a discontinuous transition of a system from a state with $N$ particles ($X^{'}_{N}$) to one with $M$ particles ($X_{M}$). Such transitions are determined not only by the configuration of the system, $X^{'}_{N}$ but depends also on the configuration of the reservoir in phase space. \redc{Ignoring the reservoir state upon collision, the change in the system state can be described in terms of a Markovian transition kernel, $K_{NM}(X^{'}_{N},X_{M})$ that is independent of time. Specifically,  $K_{NM}(X^{'}_{N},X_{M})$ is the \redc{probability that, in an infinitesimally small time interval,} the system {at $X_{M}$ makes a transition to $X^{'}_{N}$} as a result of the interaction with the reservoir. }
The probability density function, $\rho(X_{M},M,t)$, at some point $X_{M}$ of the phase space \redc{is} governed by the extended Liouville equation which we will name the Bergmann-Lebowitz Liouville equation:
\begin{eqnarray}
\frac{\partial\rho(X_{M},M,t)}{\partial t}=\{\rho(X_{M},M,t),H(X_{M})\}+\\ \nonumber
+\sum_{N=0}^{\infty}\int dX^{'}_{N}[K_{MN}(X_{M},X^{'}_{N})\rho(X^{'}_{N},N,t)-K_{NM}(X^{'}_{N},X_{M})\rho(X_{M},M,t)]
\label{liouvext}
\end{eqnarray}
where, as usual $H(X_{M})$ is the Hamiltonian of the system corresponding to the point $X_{M}$ and $\{*,*\}$ are the \redc{canonical} Poisson brackets.\\
An important point worth to mention is that the standard Liouville theorem 
\begin{equation}
\frac{d\rho(X_{M},M,t)}{dt}=0
\label{liouorig}
\end{equation}
must be replaced by a generalized Liouville theorem: 
\begin{equation}
\left[\frac{d}{dt}+\hat{Q}\right]\rho(X_{M},M,t)=f(X_{M},t)
\label{liouvext2}
\end{equation}
where 
\[
f(X_{M},t)=\sum_{N=0}^{\infty}\int dX^{'}_{N}[K_{MN}(X_{M},X^{'}_{N})\rho(X^{'}_{N},t)]
\]
and 
\[
\hat{Q}(*)=\sum_{N=0}^{\infty}\int dX^{'}_{N}[K_{NM}(X^{'}_{N},X_{M}),*]\,.
\]
\redc{The generalized Liouville theorem expresses that fact that there is a probability flux in and out of the system as a result of the interaction with the reservoir which induces the change from $N$ to $M$ particles.} The determinism of the Liouville equation, which characterizes a closed system, is now replaced by a stochastic evolution in time.\\

It is convenient to retain  the original formulation of the Liouville theorem and define an extended Liouville operator (Bergmann-Lebowitz Liouville operator):
\begin{equation}
iL^{M}_{BL}=\{*,H_{M}\}+\hat{R}(*)
\label{extop}
\end{equation}
where 
\[
\hat{R}(*)=\sum_{N=0}^{\infty}\int dX^{'}_{N}[K_{MN}(X_{M},X^{'}_{N})(*(X^{'}_{N},N,t))-K_{NM}(X^{'}_{N},X_{M})(*(X_{M},M,t))]\,.
\]
This allows us to formally write the Liouville theorem as in the standard case, namely, 
\begin{equation}
\frac{{\partial} \rho(X_{M},M,t)}{{\partial}t}+iL^{M}_{BL}\rho(X_{M},M,t)=0.
\label{berg-leb-liouv-eq}
\end{equation}
If the kernel satisfies the following integral condition \redc{(flux balance)}
\begin{equation}
\sum_{N=0}^{\infty}\int [e^{-\beta H(X^{'}_{N}){ + \beta \mu N}}K_{MN}(X_{M},X^{'}_{N})-K_{NM}(X^{'}_{N},X_{M})e^{-\beta H(X_{M}) { + \beta \mu M}}]dX^{'}_{N}=0\,,
\label{gc-cond}
\end{equation}
then the stationary Grand Ensemble is the Grand Canonical ensemble with density  
\[
\rho_{M}(X_{M},M)=\frac{1}{Q}e^{-\beta H_{M}(X_{M})+ {\beta}\mu M}\,,
\] 
with $\beta=k_{B}T$ the inverse temperature, $\mu$ the chemical potential and 
\[
Q=\sum_{M=0}^{\infty} e^{{\beta}\mu M}\int e^{-\beta H_{M}(X_{M})}dX_{M}\,.
\]
\redc{The flux balance \eqref{gc-cond} is both a necessary and sufficient condition for stationarity with respect to the Grand Canonical distribution.} 
In such a case, due to the fact that 
\[
\sum_{N=0}^{\infty}\int dX^{'}_{N}[K_{MN}(X_{M},X^{'}_{N})\rho(X^{'}_{N},N,t)-K_{NM}(X^{'}_{N},X_{M})\rho(X_{M},M,t)]=0\,,
\] 
the BL Liouville operator is formally reduced to the standard Liouvillian
\begin{equation}
iL^{M}=\{\rho(X_{M},M),H_{M}\}
\label{reducedliou}
\end{equation}
that is a Liouvillian corresponding to a Hamiltonian which propagates the system in time with variable number of particles (time-dependent, stochastically regulated). 
As a consequence the BL Liouville equation is formally reduced to the standard Liouville equation, 
\begin{equation}
\frac{\partial\rho(X_{M},M,t)}{\partial t}=\{H(X_{M}),\rho(X_{M},M,t)\},
\label{lioufin}
\end{equation}
with the number of particles being a stochastic process. 

\section{Molecular Dynamics of subsystems with varying number of molecules}
MD with varying number of particles have been developed mostly for the calculation of the excess chemical potential following the Widom insertion or the thermodynamic integration techniques \cite{widom,flor}. Such methods describe the effect of inserting or deleting a molecule in a system of $N$ molecules; they are computationally rather demanding and the calculation of the excess chemical potential is the only aim of such studies. An extension of such technique is that of hybrid MD/MC methods, in which the dynamical evolution of the MD system is interfaced with MC moves which insert or remove particles and then equilibrate the system locally before the next MD step is actuated (see discussion in Ref.\cite{flor} and references therein). Such an approach is not optimal and is computationally expensive, in fact each insertion would have costs of the order of those of Widom-like techniques for the calculation of the chemical potential.

Fully MD Grand Canonical schemes that have been developed in the past did not gain popularity due to their computational costs and a certain conceptual and theoretical artificiality. A pioneering attempt was made by Pettitt and collaborators \cite{pettit1,pettitt}; cf.~also the work of Lo and Palmer \cite{lo-palm}. The method is based on the introduction of an additional dynamic variable $s$ that represents the number of additional particles. At any instant the total number of molecules of the system can be written as $N+s$ and $s$, the new variable, corresponds to a fractional number depending on the degree of presence of an additional molecule. An extended Hamiltonian is then derived and equations of motion for $N+s$ variables are derived, moreover the knowledge {\it a priori} of excess chemical potentials is required at least when the molecular species are more than one (e.g. mixtures). It has been shown that such an approach was not optimal when applied to liquid water \cite{difpet} and further improvements were implemented in extended versions such as that of Eslami and M\"{u}ller-Plathe \cite{flor}. \redc{To our assessment, the method of Ref.\cite{flor} represents a substantial improvement of previous methods with regard to numerical robustness, nonetheless, it did not meet the expectations and the number of applications \redc{presented in literature is rather limited}.} In our view the idea of fractional \redc{particles} is conceptually very appealing, but introduces extra computational costs together with a more complex situation regarding the numerical stability of the algorithm and its implementation into pre-existing computational architectures of flexible (popular) MD codes. Later on, with the increasing success of multiscale MD techniques and the development of concurrent coupling techniques, a new generation of algorithms entered into the game \cite{annurev}. Such a category is that of adaptive molecular resolution techniques. The common idea to all methods in such a category is the definition of two main open boundary regions, one at high resolution (e.g. atomistic) and one at coarse-grained level (spherical liquid), they are interfaced by a smaller region where  molecules crossing the border acquire or loose their high resolution degrees of freedom. Molecules in the different regions are coupled via space-dependent intermolecular forces \cite{adress,jctchan,prx}, Hamiltonians \cite{ensing,kurt} or Lagrangians \cite{truh}. 
Each of these algorithms in principle can be easily converted to a Grand Canonical MD scheme if (1) the coarse-grained region is large enough to assure physically realistic particle number density fluctuations and (2) the high resolution region is large enough to be of statistical relevance (see also note in Ref.\cite{note1}).  The computational efficiency of \redc{these kinds of techniques is provably superior} to methods with varying number of particles of the previous generation (see e.g. \cite{prx,jcpan}). In this perspective they represent a realistic pathway to future MD simulations in general and in particular for those cases, in which the variation in time of the number of particles or the physics of a subsystem is of high relevance. 

In the next section we will focus on one of these techniques developed by some of the authors within the last 5-6 years with the specific aim of designing a general Grand Ensemble algorithm via adaptive resolution simulation. We will present the Grand Canonical Adaptive Resolution Simulation (GC-AdResS) method and connect its principles to the model of Bergmann and Lebowitz. In the following, the importance of such a connection for the definition and calculation of equilibrium time correlation functions  will be discussed and illustrated with numerical results.

\section{Grand Canonical-like Adaptive Resolution Simulation (GC-AdResS): Basic Principles}
\begin{figure}
   \centering
   \includegraphics[width=0.75\textwidth]{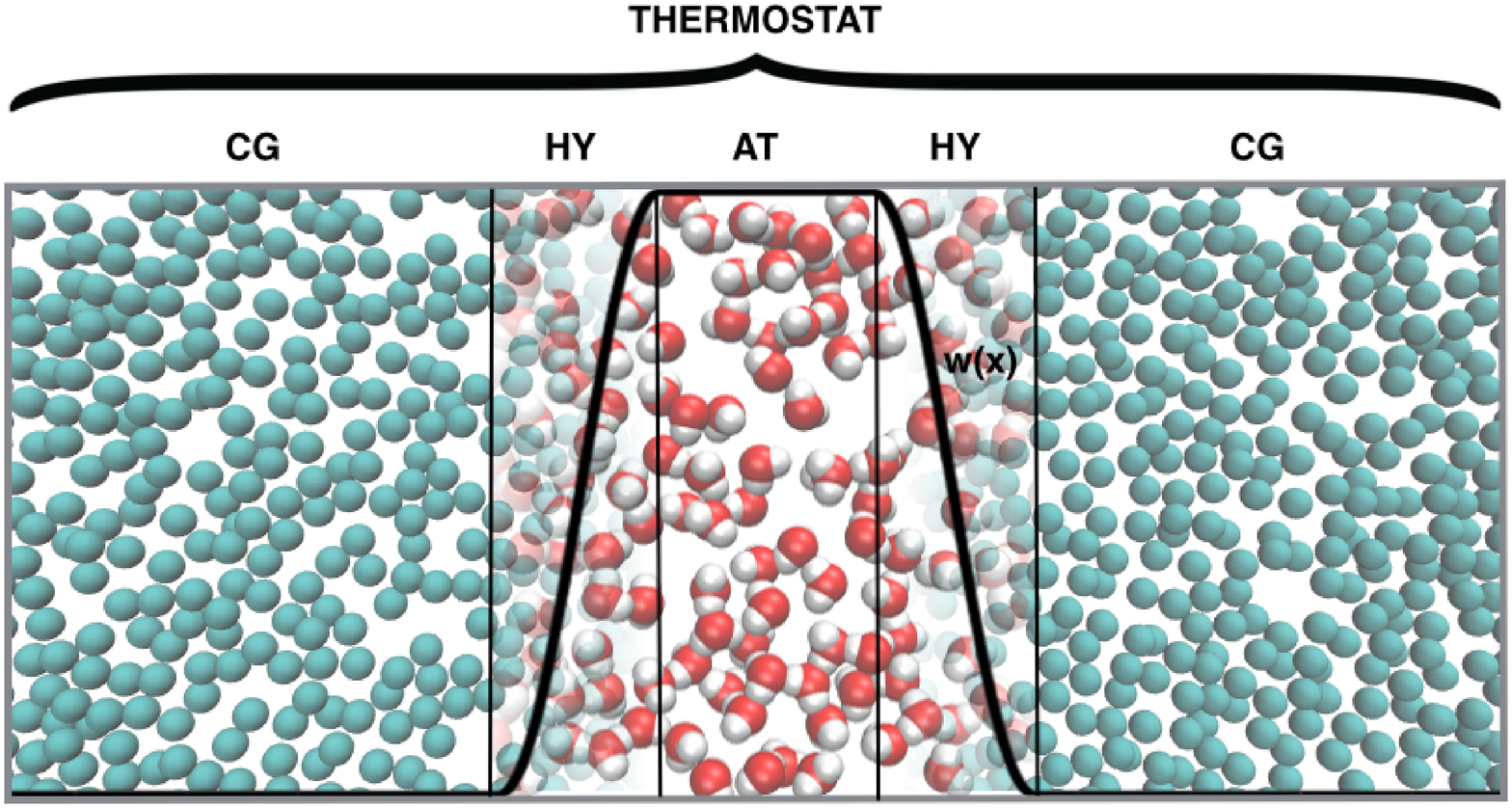}
   \includegraphics[width=0.75\textwidth]{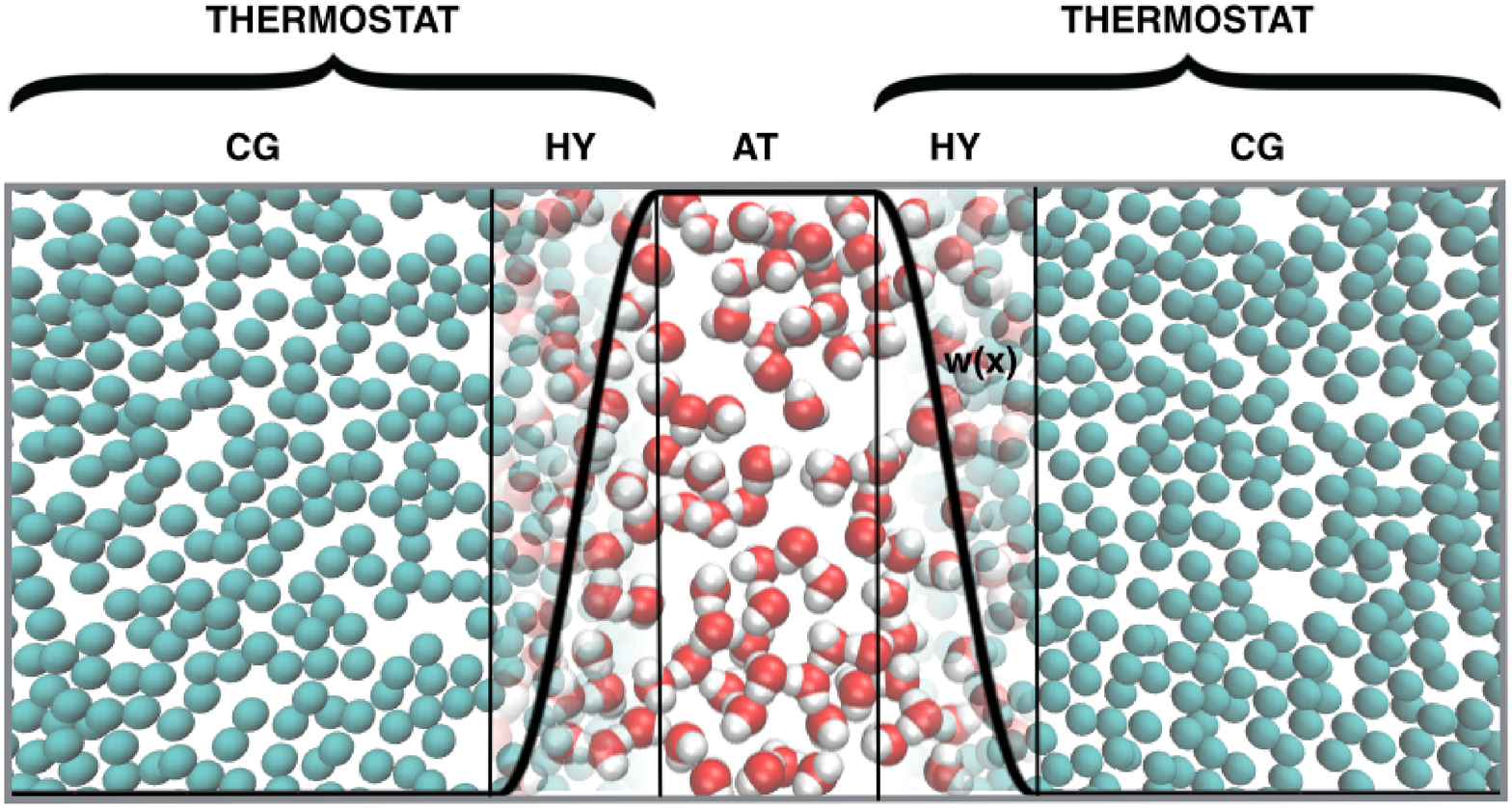}
   \caption{ Pictorial representation of the AdResS scheme; CG indicates the coarse-grained region, HY the hybrid region where atomistic and coarse-grained forces are interpolated via a space-dependent, slowly varying, function $w(x)$ and AT the atomistic region (that is the region of interest). Top, the standard \redc{set-up} with the thermostat that acts globally on the whole system. Bottom, the ``local'' thermostat technique employed in this work. }
   \label{fig1}
 \end{figure}

\redc{The basic structure of the original AdResS \cite{adress} is based on an intuitive technical requirement, namely, the construction of a numerical scheme which allows the system to pass smoothly from an atomistic to a coarse-grained dynamic evolution in space in such a way that the dynamics of the atomistic part is not perturbed significantly by the dynamics of the coarse-grained part and vice versa. The flow of molecules between the two regions must constructed in such a way that the exchange happens under conditions of thermodynamic equilibrium; it is expected that static and dynamical properties of the atomistic region must be the same as in an equivalent subsystem of a fully atomistic reference simulation. The construction of such a numerical machinery is reported step by step below}:
\begin{itemize}
\item The space is partitioned in three regions, one characterized by atomistic resolution (AT) and one characterized by  coarse-grained (usually spherical) resolution (CG) and a relatively small interface region with hybrid resolution (transition region or hybrid region) ($\Delta$ or HY).
\item Molecules in the different regions are smoothly coupled through a spatial interpolation formula for the forces:
\begin{equation}
{\vect F}_{i,j}=w(\vect r_i)w(\vect r_j){\vect
  F}_{i,j}^{\AT}+[1-w(\vect r_i)w(\vect r_j)]{\vect F}^{\CG}_{i,j} 
\label{eqforce}
\end{equation}
where $i$ and $j$ indicates two molecules, ${\vect F}^{\AT}$ is the
force corresponding the atomistic interactions ($U_{\AT}$) (e.g. standard Lennard-Jones or Coulomb atomistic potential) and  ${\vect F}^{\CG}$
is the force corresponding to \redc{the coarse-grained interaction potential $U_{\CG}$} (e.g. standard COM-COM potential, where COM stays for ``the center of mass''), $\vect r$ is the COM position of
the molecule and $w(x)$ is a smooth function, defined over the transition region ($\Delta$), which goes from $0$
to $1$ (or vice versa). It acts in such a way that the lower resolution is 
slowly transformed in the high resolution (or vice versa),
as illustrated in Fig.\ref{fig1}.
\item A thermodynamic force, defined via first principles of thermodynamics, acts on the COM of each molecule and a thermostat is added to assure the overall thermodynamic equilibrium at the chosen temperature. The thermodynamic force is derived in such a way that:
  $p_{\AT}+\rho_{0}\int_{\Delta}{\vect F}_{\thf}({\vect r})d{\vect r}=p_{\CG}$,
  where $p_{\AT}$  is the chosen pressure of the atomistic system (region), $p_{\CG}$ is the pressure of the coarse-grained model, $\rho_{0}$ is the chosen molecular density of the atomistic system (region)~\cite{prl12} \redc{(the \redc{explicit} expression of ${\vect F}_{\thf}({\vect r})$ will be specified later on)}. A thermostat is added to take care of the loss/gain of energy in the transition region. This is the first step to pass from the original intuitive idea of AdResS to a well founded Grand Canonical framework of the method. In the original AdResS \redc{set-up}, the thermostat acts over the whole system (see top panel of Fig.\ref{fig1}), in this work the idea has been developed further and in order to match the requirements of the reservoir of the BL model \redc{for the calculation of equilibrium time correlation functions}, we have constructed a \redc{set-up}, in which the thermostat is applied to the reservoir only (i.e.~hybrid and coarse-grained region); see bottom panel of Fig.\ref{fig1}.
\end{itemize}
In Ref.\cite{prx} and in Ref.\cite{jctchan} necessary conditions in $\HY$ were \redc{derived so that the spatial probability distribution in the atomistic region was close to that of a fully atomistic reference system up to a certain chosen order. The probability distribution is that of a Grand Canonical ensemble, hence the name \emph{Grand-Canonical-AdResS (GC-AdResS)}. We define the $m$-th order statistics of a joint probability distribution of $M$ molecules,   $p({\vect r}_{1}, \cdots, {\vect r}_{M})$, as}
\begin{equation}
p^{(m)}({\vect r}_{1}, \cdots, {\vect r}_{m})=\int p({\vect r}_{1}, \cdots, {\vect r}_{m},{\vect r}_{m+1}, \cdots, {\vect r}_{M}) \: d{\vect r}_{m+1}\cdots d{\vect r}_{N}\,.
\label{order}
\end{equation}
The molecular number density $\rho({\vect r})$ corresponds to the first order, the radial distribution function to the second, three-body distributions to the third order \redc{statistics and so on; examples of how the statistics in the atomistic region is reproduced will be shown later on.
We emphasize that, by construction of the method, the accuracy in the atomistic region is independent of the accuracy of the coarse-grained model, thus, in the coarse-grained region, one can use a generic liquid of spheres whose only requirement is that it has the same molecular density of the reference system (i.e. we need only to know the distribution of the reservoir and not its microscopic details, which is in accordance with the basic principle of construction of the BL reservoir)}. \redc{It was numerically \redc{demonstrated} for the case of liquid water that the target Grand Canonical distribution, numerically defined as the probability distribution of a subsystem (of the size of the atomistic region in GC-AdResS) in a large\redc{, fully} atomistic simulation, is accurately reproduced \redc{to (at least) third order.}} To complete the idea of Grand Canonical-like \redc{set-up}, it was shown that the sum of the works of ${\vect F}_{th}({\vect r})$ and the thermostat in the transition region is equivalent to the difference of the chemical potentials between the atomistic and coarse-grained resolution (at the given thermodynamic conditions). Details will be given later on.

\begin{figure}
   \centering
   \includegraphics[width=0.75\textwidth]{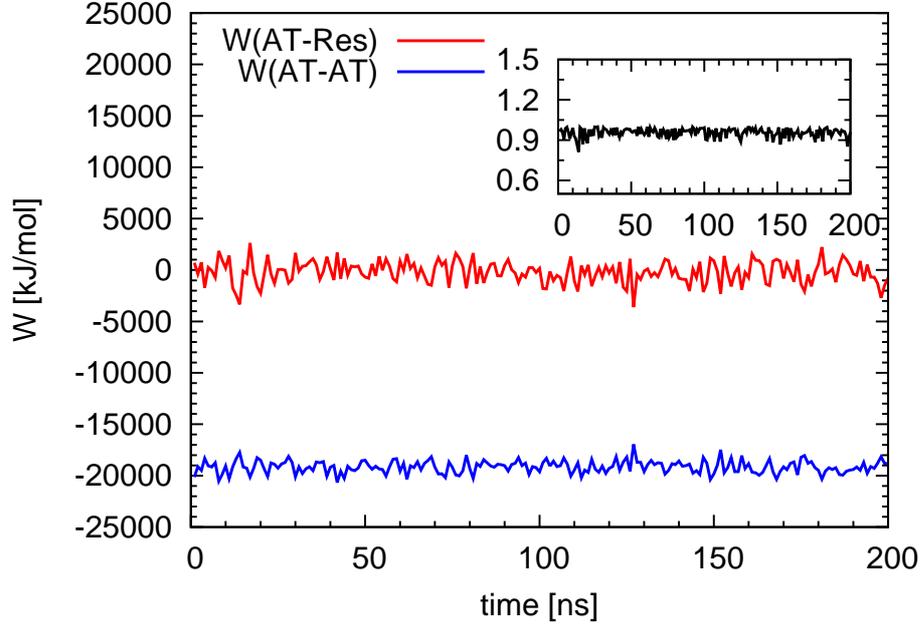}
   \caption{ \redc{Main figure: Potential energy of the subsystem only as a function of time, $W_{AT-AT}(t)$ compared to the energy associated to the interaction between subsystem and reservoir, $W_{AT-RES}(t)$; the former is at least one order of magnitude than the latter. Inset: The relative effect of the interaction between the AT region and the reservoir as a function of time : $\frac{|W_{AT-AT}(t)|-|W_{AT-RES}(t)|}{|W_{AT-AT}(t)|}$, it can be clearly seen that the contribution is, at most, of $10\%$. It must be underlined that in a test done with a much larger \redc{system, the} effect goes below $1.0 \%$.}}
   \label{wwuo}
 \end{figure}

\redc{The construction of a thermostat that acts only in the hybrid and CG regions makes the reservoir of GC-AdResS the effective technical translation of the reservoir hypothesized by Bergmann and Lebowitz in their model}. A detailed discussion of the validity of the approximations of the method in the light of the theoretical hypothesis of the BL model is outlined in the next section.
\redc{
\section{Bergmann-Lebowitz model and  GC-AdResS}
In this section we analyze the correspondence between the BL model and GC-AdResS, more specifically we will discuss the possible mathematical mapping between the formulas of the two models and analyze the corresponding algorithmic meaning.
\subsection{Mapping the Hamiltonian of the AT region}
For the $i$-th molecule, at position, $\vect r_i$ in the AT region of AdResS (hereafter named ``system''), we have $w(\vect r_i)=1$, thus the corresponding force can be divided in two contributions; one is the force generated by the interaction of molecule $i$ with molecules of the AT region:
\begin{equation}
{\vect F}_{i,j}={\vect
  F}_{i,j}^{\AT} , \forall j\in \AT
\label{eqforceat}
\end{equation}
and one is the force generated by the interaction with molecules of the reservoir
\begin{equation}
{\vect F}_{i,j}=w(\vect r_j){\vect
  F}_{i,j}^{\AT}+[1-w(\vect r_j)]{\vect F}^{\CG}_{i,j} , \forall j\in \HY+CG.
\label{eqforcehyb}
\end{equation}
Eq.\ref{eqforceat} implies the possibility of expressing the force acting on molecule $i$ in terms of the gradient of the atomistic potential:
\begin{equation}
{\vect F}_{i}=\sum_{j\neq i}{\vect F}_{i,j}^{\AT}=\sum_{j\neq i}\nabla_{i}U_{\AT}
\label{grad1}
\end{equation}
where $\nabla_{i}$ is the gradient w.r.t. molecule $i$.
Eq.\ref{eqforcehyb} expresses instead the action of molecules of the reservoir on molecule $i$, that is an external force. 
The system-reservoir coupling term of  Eq.\ref{eqforcehyb} \redc{rules out} the existence of a microscopic Hamiltonian for the system (embedded in the reservoir) and thus impedes a straightforward correspondence between the BL Hamiltonian, $H_{M}$, of Eq.\ref{reducedliou} (or $H(X_{M})$ of Eq.\ref{lioufin}) and the Hamiltonian of the AT region, $H_{AT}$, of the AdResS model. 
However here we want to advocate the view that the AdResS model can be mapped to the BL framework, even though a rigorous derivation of the BL kernel from a microscopic model is beyond the scope of this paper. We will provide numerical evidence for this point of view later on in the text. Roughly speaking, one may argue that the non-integrable part of the dynamics in the HY region represents a boundary effect that can be absorbed in the definition of the transition kernel. To elaborate on this point, we first notice that Eq.\ref{eqforcehyb} can recast as:
\begin{equation}
{\vect F}_{i}=\sum_{j\in \HY+\CG}[w(\vect r_j){\vect
  F}_{i,j}^{\AT}+[1-w(\vect r_j)]{\vect F}^{\CG}_{i,j}]=\sum_{j\in\HY+\CG}[w(\vect r_j)\nabla_{i}U_{AT}+[1-w(\vect r_j)]\nabla_{i}U_{CG}].
\label{grad2}
\end{equation}
Hence the net force on the $i$-th particle can be considered as a (non-local) gradient field that is instantaneously generated by the external field generated by the other molecules. As a consequence, the energy of the $i$-th molecule at time $t>0$ associated with the coupling force of Eq.\ref{grad2} can be defined as 
\begin{equation}
W^{i}_{AT-RES}(t)=\sum_{j\in\HY+\CG}[w(\vect r_j)U^{ij}_{AT}+[1-w(\vect r_j)]U^{ij}_{CG}], 
\label{waresat}
\end{equation}
where the $U_{\cdot}^{ij}$ represent the interaction energies between molecule $i$ at position ${\bf r}_{i}$ and the other molecules sitting at ${\bf r}_{j}$. The  
 total energy in the system at time $t$ is then defined as
\begin{equation}
W_{AT-RES}(t)=\sum_{i\in AT}W^{i}_{AT-RES}(t)\,.
\label{toten}
\end{equation}
The quantity of Eq.\ref{toten} should be compared to the amount of energy, $W_{AT-AT}$, corresponding to the interaction between molecules of the AT region only: $W_{AT-AT}(t)=\sum_{i<j}U^{ij}_{AT}; i,j\in AT$.
If 
\begin{equation}
\frac{|W_{AT-AT}(t)|-|W_{AT-RES}(t)|}{|W_{AT-AT}(t)|}\approx 1; \forall t
\label{criter}
\end{equation}
then it seems reasonable to approximate the total energy of the atomistic system by the Hamiltonian of the AT region, 
\begin{equation}
H_{AT}\approx H_{AT-AT}\,.
\label{finham}
\end{equation}
which corresponds to the microscopic Hamiltonian $H_{M}$ of the BL model.
For all practical purposes, Eq.\ref{criter} holds true when the HY region can be considered thin compared to the AT region and when the AT region is large. In this case, given the typical cut off radius of interactions across the HY region, there is no direct interactions between the AT region and the CG region.
However, Eq.\ref{criter} may not hold under more realistic conditions as they are routinely used in AdResS simulation, with a not too large AT region and an HY region that is not too thin so as to avoid numerically stiff systems. Fig.~\ref{wwuo} displays the behaviour of $W_{AT-AT}(t)$ and $W_{AT-RES}(t)$ for a system of 5000 molecules (about 450 in the AT region) that represents a worst case scenario in this regard. We observe that $W_{AT-AT}(t)$ is at least one order of magnitude larger than $W_{AT-RES}(t)$, so that the modeling error in terms of equilibrium expectation values that arises from replacing $H_{M}$ of the BL model by $H_{AT-AT}$ is about $10\%$. This estimate is clearly an upper bound for the model error and the neglected terms can be remodeled by an appropriate choice or parametrization of the kernel, as will be discussed in the next paragraph. A numerical test with a system close to the ideal condition of thermodynamic limit (100000 molecules, with 20000 in the AT region) shows that the \redc{energy contribution $W_{AT-RES}(t)$ is less than $1\%$. Hence, for all practical pusposes, $H_{M}=H_{AT-AT}$ fully specifies the microscopic characteristics of the AT system.}

\subsection{The action of the reservoir and the interpretation of the transition kernel}
We shall proceed with discussing the correspondence between the BL and GC-AdResS reservoirs and the role of the kernel. To this end we recall that, in the BL framework, $K_{NM}(X^{'}_{N},X_{M})$ is the transition rate for the system in state $X_{M}$ to make a transition to $X^{'}_{N}$ as a result of the interaction with the reservoir. Further recall that (\ref{gc-cond}) is both necessary and sufficient for the system to admit a unique stationary grand canonical distribution. This implies that (\ref{gc-cond}) holds by construction of GC-AdResS that is ergodic with respect to the grand canonical distribution. This clearly does not uniquely determine the transition kernel, nor does it guarantee its existence, but we will discuss how the transition kernel can be interpreted within the GC-AdResS \redc{framework.

The} influence of the GC-AdResS reservoir on the dynamics in the AT region comprises three contributions: (a) the thermostat, (b) the thermodynamic force, and, (c), the coupling force (\ref{grad2}). Firstly, the function of the thermostat is that of assuring thermal stability of the reservoir and, as a consequence, of the system. \redc{Thermal stability is guaranteed by irreducibility of the kernel, so that it is possible to go from any region of the AT phase space to any other region with a positive probability \cite{MeynTweedie}. A slightly stronger condition is that the dynamics are ergodic which is guaranteed by the recurrence of the dynamics, i.e., every phase space region is visited infinitely often with a positive probability. \redc{We should emphasize that this condition is known to be false for almost all deterministic Hamiltonian systems expect for certain billiards and geodesic flows on surfaces of constant negative mean curvature, therefore we use a gentle stochastic thermostat in AdResS.} We refrain from going into details here and instead refer to \cite{TheilLeimkuhler2009} for a discussion of this issue.} \\
Secondly, the thermodynamic force, is computed via the following iterative procedure: 
\begin{equation}
F_{k+1}^{th}(x)=F_{k}^{th}(x) - \frac{M_{\alpha}}{[\rho_{o}]^2\kappa}\nabla\rho_{k}(x)\,.
\end{equation}
The fixed point iteration converges locally as the density profile across the HY region becomes flat. This requires an exchange of particles between the AT and the GG regions, hence the thermodynamic force has the effect that the number of particles in the AT region vary in such a way that the average number density is constant (equal to the fixed target density).
This also means that\redc{, by transporting the action of the thermostat, the effect of $F_{th}(x)$ is to impose} the stationary distribution of the reservoir at the first order ($\rho(x)$), independently of the interaction between the reservoir and the system; this condition is equivalent to the main condition requested/satisfied by the reservoir in the BL model.
The computation of the thermodynamic force corresponds to the equilibration procedure of GC-AdResS; once the fixed-point iteration has converged (which it does at least locally), the thus obtained force is used for the simulations of production runs. The chemical potential, $\mu=\mu_{\AT}$, in \eqref{gc-cond} is then automatically determined according to the equation (see \cite{prx,jcpan} for details)
\begin{equation}
\mu_{\CG}=\mu_{\AT}+\omega_{\thf}+\omega_{\thermo}\,,
\label{mu}
\end{equation}
where $\omega_{\thf}=\int_{\HY}{\vect F}_{\thf}({\vect r})d{\vect r}$ and $\omega_{\thermo}=\int_{\HY}\nabla w(\vect r)\langle w( U^{\AT}-U^{\CG})\rangle_{\vect r} d{\vect r}+\omega_{gas}$, $w(\vect r)$ is the force interpolation function of Eq.\ref{eqforce} and $\omega_{gas}$ is the chemical potential in absence of intermolecular interactions and $\langle\cdot\rangle_{\vect r}$ indicates the conditional equilibrium average for fixed AT configurations.\\
Eq.~\eqref{mu} is the minimal necessary condition that the  GC-AdResS system  should satisfy in order to have a Grand-Canonical like molecular dynamics, i.e.~to satisfy the condition Eq.~\eqref{gc-cond}, and, as stated above it is imposed by the thermodynamic force. 
The numerical verification that indeed the AT region of GC-AdResS  behaves as a Grand Canonical ensemble is then made by comparing quantities calculated in the GC-AdResS AT system with those calculated in an equivalent subsystem of a \redc{fully} atomistic reference system (see results in section \ref{statprop}). A subsystem in a \redc{fully} atomistic simulation, if the subsystem and the total system are large enough,  is a natural Grand Canonical system. It follows that if the reservoir in the fully atomistic reference system and the GC-AdResS reservoir have the identical insertion/deletion behaviour (Eq.~\eqref{gc-cond}), they must spend the same amount of energy in insertion/deletion, i.e.~have the same chemical potential difference between the AT region and the rest of the system.
This implies that the condition of Eq.~\eqref{gc-cond} in the BL model corresponds to Eq.\ref{mu} of the GC-AdResS model.\\

Thirdly, in accordance with the above reasoning, the coupling force in \eqref{grad2} does not give a major energetic contribution to the AT interactions. Nevertheless it involves strong repulsive forces that prevent the molecules entering in the AT region from overlapping with molecules that are already in the AT region, which would produce (numerical) singularities that would automatically stop the simulation. This soft collision-avoidance has the effect that the smooth density of the transition kernel is exponentially decaying outside the admissible (non-overlapping) particle configurations. Hence, even though the coupling force can be conceptually neglected as far as the construction of the transition kernel is concerned, it plays a key role in the numerical simulation as it imposes collision-avoidance between AT and HY/CG particles in a robust and numerically efficient way.\\
Altogether, even though we cannot give a rigorous derivation of the BL  kernel within the GC-AdResS framework, we have described how some of the properties of the kernel that guarantee well-posedness of the dynamics can be inferred from the properties of the various force contributions. It is unclear whether it is possible to write the kernel explicitly in terms of the forces. We shall argue that, even though such a direct link may not exist, it is still possible to realize the BL model numerically, and CG-AdResS does exactly this. For example, stochastic insertion/removal of molecules in the system (cf.~\cite{pettit1,flor}) can be used to realize $K_{NM}(X^{'}_{N},X_{M})$ in a Monte Carlo fashion. The basic idea is that a molecule is inserted in the system by searching a location that is close to a minimum free energy configuration  followed by a local equilibration where the rate of insertion is defined by the chemical potential of the system in accordance with \eqref{gc-cond}; equivalently\redc{,} within the framework of GC-AdResS the random particle number fluctuations (in the AT region) are realized by the self-consistent  iteration of the thermodynamic force.

\subsection{Bergmann-Lebowitz model as conceptual guideline for the calculation of equilibrium time correlation functions in the GC-AdResS}
}
According to popular textbooks of statistical mechanics and molecular simulation (see e.g. \cite{tuck}), the general definition of the equilibrium time correlation function, $C_{AB}(t)$ between two physical observables, $A$ and $B$ is:
\begin{equation}
\begin{aligned}  
C_{AB}(t) =\langle a(0)b(t)\rangle &  =\int d{\bf p}d{\bf q}f({\bf p},{\bf q}) a({\bf p},{\bf q})e^{iL_{t}}b({\bf p},{\bf q})\\
& =\int d{\bf p}d{\bf q}f({\bf p},{\bf q}) a({\bf p},{\bf q})b({\bf p}_{t}({\bf p},{\bf q}),{\bf q}_{t}({\bf p},{\bf q}))
\end{aligned}
\label{eq1}
\end{equation}
where, $a({\bf p},{\bf q})$ and $b({\bf p},{\bf q})$ are phase space functions corresponding to the observables $A$ and $B$ respectively, $a(0)=a(t=0)$ and $b(t)$ is the function at time $t$, $f({\bf p},{\bf q})$ is the equilibrium distribution function and the dynamics is generated by the Liouville operator $iL$. The \redc{notation} ${\bf p}_{t}({\bf p},{\bf q}),{\bf q}_{t}({\bf p},{\bf q})$ is taken from Ref.\cite{tuck} and indicates the time evolution at time $t$ of the momenta and positions with initial condition ${\bf p},{\bf q}$. For a canonical ensemble the definition in Eq.\ref{eq1} takes the explicit form:
\begin{equation}
C_{AB}(t)=\frac{1}{Q_{N}}\int d{\bf p}d{\bf q} e^{-\frac{H_{N}({\bf p},{\bf q})}{kT}}a({\bf p},{\bf q}) b({\bf p}_{t}({\bf p},{\bf q}),{\bf q}_{t}({\bf p},{\bf q})).
\label{eq2}
\end{equation}
where $Q_{N}$ is the Canonical partition function and $H_{N}({\bf p},{\bf q})$ the Hamiltonian of a system with $N$ (constant) molecules.
According to Eq.\ref{eq2}, the numerical calculation of $C_{AB}(t)$ can be done by calculating $a({\bf p},{\bf q})$ and $b({\bf p}_{t}({\bf p},{\bf q}),{\bf q}_{t}({\bf p},{\bf q}))$ along each MD trajectory and averaging over all the data obtained. The trajectories must be long enough so that the basic requirements of ergodicity and statistical relevance of the data can be safely assumed. 
In such a case the dynamics generated by the Liouvillian operator is well defined, since the Liouville operator is well defined by the Hamiltonian of $N$ molecules:
\begin{equation}
iL=\sum_{j=1}^{N}\left[\frac{\partial H}{\partial{\bf p}_{j}}\frac{\partial}{{\partial{\bf q}^{j}}}-\frac{\partial H}{\partial{\bf q}^{j}}\frac{\partial}{{\partial{\bf p}_{j}}}\right]=\left\{*,H\right\}
\label{lioufix}
\end{equation}
Now let us \redc{formally generalize} Eq.\ref{eq2} to the case of a Grand Canonical ensemble:
\begin{equation}
C_{AB}(t)=\frac{1}{Q_{GC}}\sum_{N}\int d{\bf p}_{N}d{\bf q}_{N} e^{-\frac{[H_{N}({\bf p}_{N},{\bf q}_{N})-\mu N]}{kT}}a({\bf p}_{N},{\bf q}_{N}) b({\bf p}_{t}({\bf p}_{N},{\bf q}_{N}),{\bf q}_{t}({\bf p}_{N},{\bf q}_{N})).
\label{eq3}
\end{equation}
where $Q_{GC}$ is the Grand-Canonical Partition function, $\mu$ the chemical potential and $N$ the number of particles (now varying in time) of the system.
\redc{The difficulty lies in how to interpret the quantity} $b({\bf p}_{t}({\bf p}_{N},{\bf q}_{N}),{\bf q}_{t}({\bf p}_{N},{\bf q}_{N}))$. In fact at a given time $t$ the system evolved from its initial condition and it is likely to have a number of particles/molecules $N^{'}$ different from the initial state.
\redc{The correspondence of GC-AdResS with the model of Bergmann and Lebowitz plays a key role for making sense of $b({\bf p}_{t}({\bf p}_{N},{\bf q}_{N}),{\bf q}_{t}({\bf p}_{N},{\bf q}_{N}))$ in the numerical simulation as Eq.\ref{reducedliou} states that there exists a Liouvillian $iL^{M}$, the action of which is to evolve the system from $({\bf p}_{N},{\bf q}_{N})$ to $({\bf p}_{t},{\bf q}_{t})$ with $N^{'}$ molecules. As we have argued, the operator $iL^{M}$ is well defined within the  GC-AdResS framework.} 
Thus the correspondence between the BL model and GC-AdResS leads to the following \redc{ready-to-use definition of the equilibrium time correlation functions for numerical simulations with CG-AdResS: {\it ``if a molecule leaves the AT region in the observation time window, its contribution to the correlation function is neglected''}. This principle is in agreement with the philosophy of the BL model, which asserts that a molecule entering into the reservoir loses its microscopic identity.}

\section{Numerical Results}
Here we report numerical results for liquid water (SPC/E model) at room conditions\redc{. The Section is divided in two parts:} The first is dedicated to the calculation of static properties with the intention of \redc{demonstrating---numerically---}that GC-AdResS produces results typical of a natural Grand Canonical system (as defined before). The second part is dedicated to the calculation of the equilibrium time correlation functions. In such a case the exchange of particles with the reservoir poses\redc{,} on the one hand\redc{,} the conceptual question of how to define the Liuoville operator of the atomistic region and\redc{,} on the other hand\redc{,} the practical question of how to count correlations when a molecules leaves the atomistic region or enters in it. The theoretical concepts of section II actually give the guidelines to solve both problems. We will first prove that with the definitions taken from section II GC-AdResS gives the same results as those of an open subsystem of a \redc{fully} atomistic NVE simulation. Next, since in the thermodynamic limit all ensembles are equivalent, we expect, for physical consistency, that by increasing the size of the atomistic region\redc{,} results systematically converge to those of a full NVE simulation where the calculations are performed over the whole system; the numerical results reported below \redc{confirm our expectations}.
\subsection{static properties}
\label{statprop}
\begin{figure}
   \centering
   \includegraphics[width=0.5\textwidth]{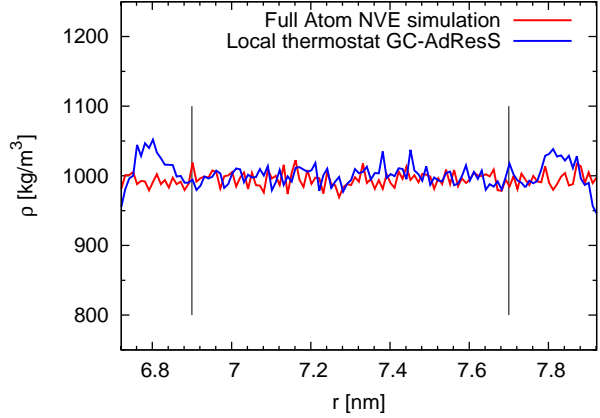}
   \caption{ Molecular number density calculated with AdResS where the thermostat is acting only in the reservoir. Results are compared with the density obtained for an equivalent subsystem ($1.2 nm$) in a full atomistic NVE simulation. A discrepancy of about $5\%$ can be observed at the border of the AT region (vertical lines). Besides the fact that a discrepancy of $5\%$ is not dramatic, in general the rigorous application of GC-AdResS requires that part of the hybrid region contains a buffer of \redc{fully} atomistic molecules. Here we want to show that even in the worst-scenario-case\redc{,} the numerical accuracy is still very high.}
 \label{rho}
 \end{figure}

\begin{figure}
   \centering
   \includegraphics[width=0.5\textwidth]{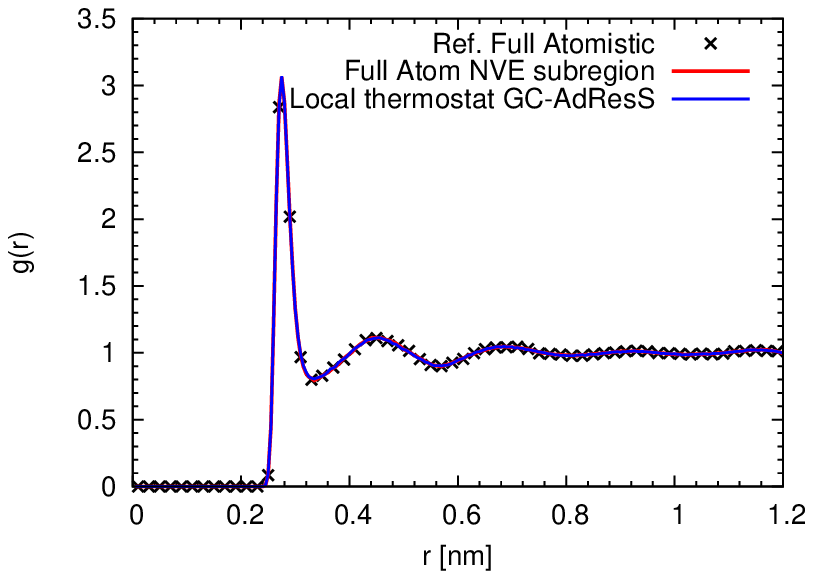}
   \includegraphics[width=0.5\textwidth]{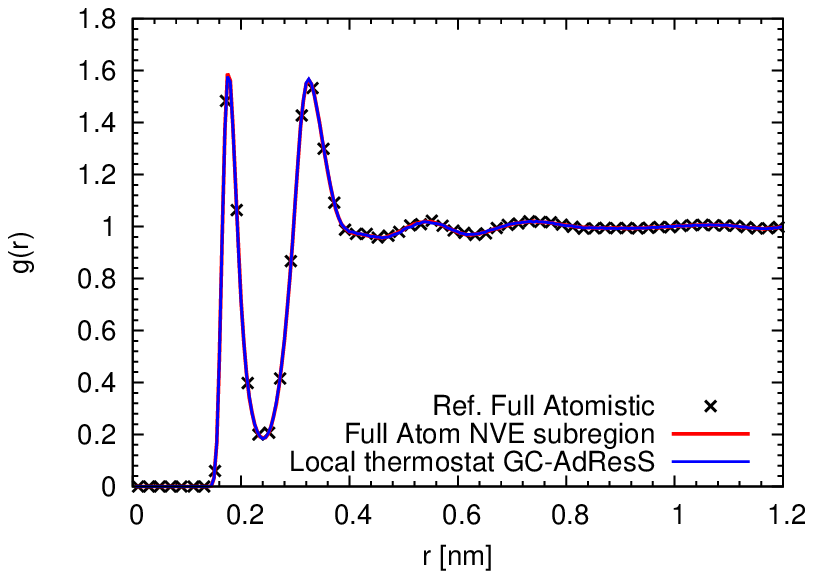}
   \includegraphics[width=0.5\textwidth]{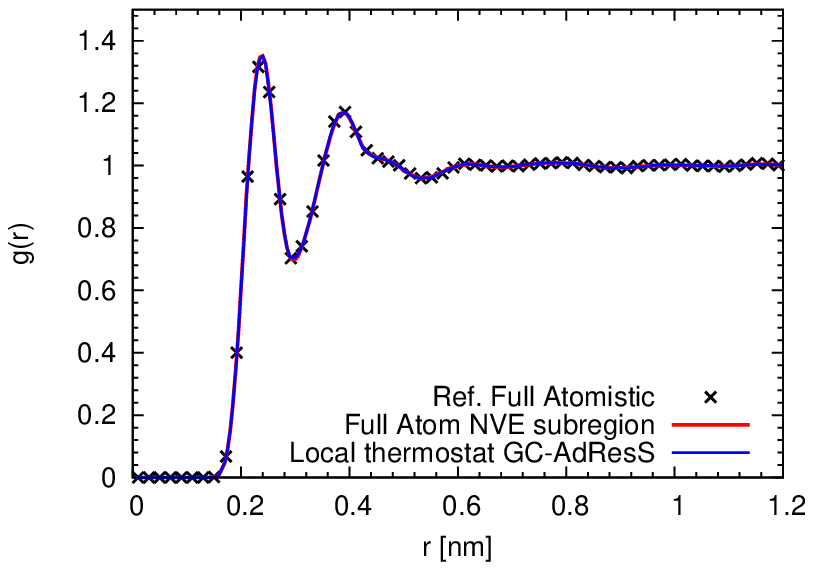}
   \caption{ Oxygen-oxygen (top), oxygen-hydrogen (middle) and hydrogen-hydrogen (bottom) radial distribution functions calculated with AdResS where the thermostat is acting only in the reservoir. Such functions are compared with the results obtained for an equivalent subsystem ($1.2 nm$) in a \redc{fully} atomistic NVE simulation \redc{and with the same quantity calculated over the entire system in the \redc{fully} atomistic simulation; the agreement is highly satisfactory.}}
   \label{gr}
 \end{figure}

\begin{figure}
   \centering
   \includegraphics[width=0.5\textwidth]{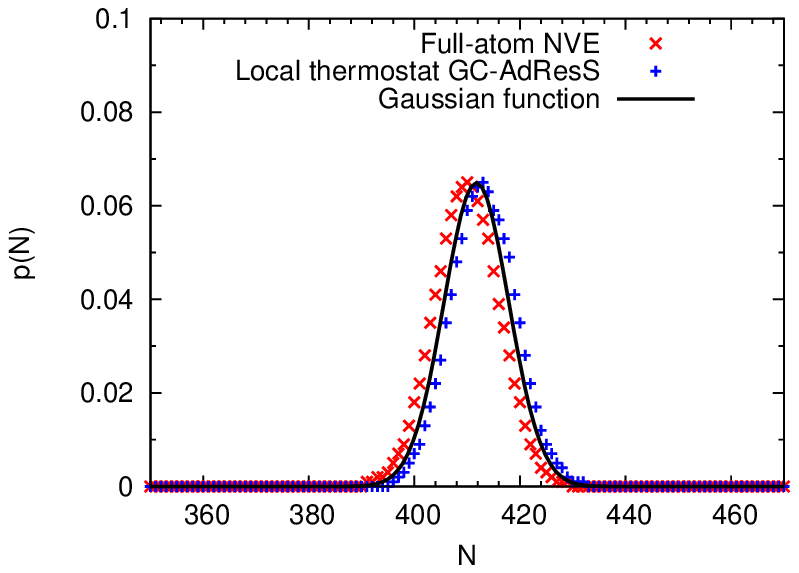}
   \includegraphics[width=0.5\textwidth]{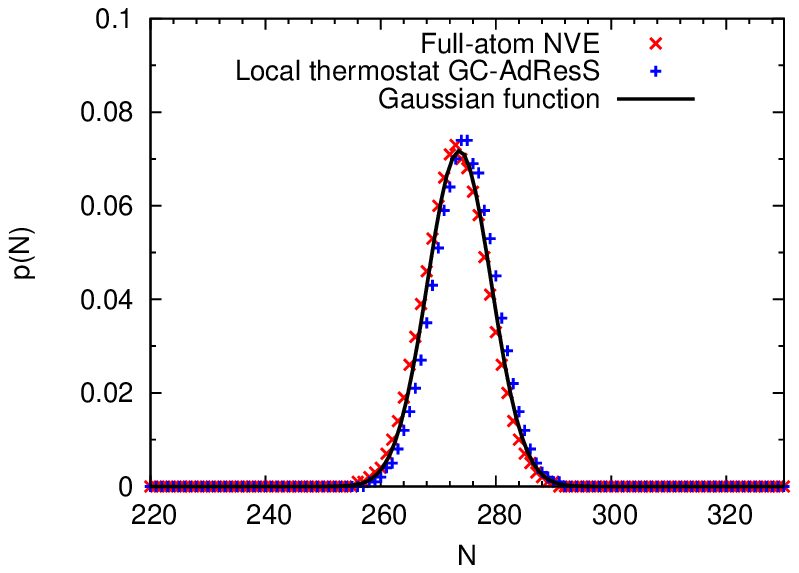}
   \caption{(top) Particle number probability distribution of AdResS compared with the equivalent NVE subsystem. 
 The subsystem employed for this calculation is an open subsystem embedded into the  NVE global system (i.e. we consider only molecules in a subregion of the global NVE system). Such a subsystem has the same size of the atomistic region of AdResS; it freely exchanges molecules with the rest of the system.
 The shape of both curves is a Gaussian (reference black continuous curve); the curve of AdResS is shifted compared to the NVE results of only of \textbf{two particles}. However, if we consider the additional atomistic buffer (bottom), as it should be if the principles of GC-AdResS are rigorously applied, then the two curves overlap.}
   \label{pn}
 \end{figure}

Figs.\ref{rho},\ref{gr},\ref{pn} and Tables \ref{tablecp1} and \ref{tablecp2} show static properties calculated with local thermostat GC-AdResS compared to NVE full atomistic calculations of an equivalent subsystem. 
In particular Figs.\ref{rho},\ref{gr} show that GC-AdResS, with the current definition of reservoir, can properly reproduce the probability distribution of a natural Grand Canonical at least up \redc{to second} order.
The difference with results of Ref.\cite{prx} is that the transition region is considerable smaller and that the thermostat acts only in the reservoir. \redc{A few remarks in this regard are in order:} In Fig.\ref{rho} the number particle density of GC-AdResS agrees in a satisfactory way with that of the NVE calculation, the largest deviation (below $5\%$) is at the border of the atomistic region with the hybrid region. This is due to the abrupt absence of the thermostat. The effect is anyway negligible, however\redc{,} there are three technical options which allow to make the effects of such difference even smaller: (a)  apply the rigorous GC-AdResS protocol and consider an additional (but, differently from Ref.\cite{prx}, negligible) atomistic buffer as part of the transition region, (b) require that the convergence of the thermodynamic force is stricter, (c) slowly switch off the thermostat in the transition region near the atomistic region. Here we have opted for the simpler option (a),
because in any case the effects of this discrepancy on the calculation of physical quantities produce no more than $10\%$ of deviation compared to the reference data (see discussion below).  
\redc{Fig.\ref{pn} reports} the particle number probability distribution of the subsystem compared with an equivalent NVE subsystem, the shape of both curves is a Gaussian and the curve of GC-AdResS is indeed shifted compared to the NVE of reference, but \redc{only for two two particles and only very little}. If we apply the rigorous GC-AdResS protocol and consider an additional (negligible) atomistic buffer, then the two curves essentially overlap, see Fig.\ref{pn} (bottom). 
 Table \ref{tablecp1} shows the robustness of the method as a Grand Canonical \redc{set-up} for the calculation a thermodynamic property, that is  energy fluctuation and the covariance (see Appendix for definitions and technical details).
Regarding the accuracy, in the worst case the deviation is no more than $10\%$, which would be already numerically satisfactory. However if we  apply the rigorous GC-AdResS protocol (as in  Fig.\ref{pn} (bottom) ) the maximum deviation falls down to $3\%$ only, see Table \ref{tablecp2}. 
\begin{table}[htpb]
\begin{center}
\begin{tabular}{ccc}
\hline \hline
 Quantity & Full-Atomistic & GC-AdResS  \\
\hline
$\frac{\langle E^{2} \rangle - \langle E \rangle^{2}}{\langle E \rangle}$  & $20.6 \pm 0.4$  & $19.3 \pm 0.4$  \\
$\frac{\langle NE \rangle - \langle N \rangle  \langle E \rangle}{\langle N \rangle}$  & $4.4 \pm 0.2$  & $3.9 \pm 0.2$  \\
\hline \hline
\end{tabular}
\caption{Thermodynamic fluctuations calculated in atomistic subregion ($EX$=1.2) in GC-AdResS and full-atom simulations. There is a discrepancy of around 5-10\% between the results of GC-AdResS and those of the reference full-atom simulation. }
\label{tablecp1}
\end{center}
\end{table}

\begin{table}[htpb]
\begin{center}
\begin{tabular}{ccc}
\hline \hline
 Quantity & Full-Atomistic & GC-AdResS  \\
\hline
$\frac{\langle E^{2} \rangle - \langle E \rangle^{2}}{\langle E \rangle}$  & $27.1 \pm 0.5$  & $26.4 \pm 0.5$  \\
$\frac{\langle NE \rangle - \langle N \rangle  \langle E \rangle}{\langle N \rangle}$  & $5.1 \pm 0.2$  & $4.9 \pm 0.2$  \\
\hline \hline
\end{tabular}
\caption{Same quantities as above calculated in the region excluding the (negligible) part where the density is $5\%$ off compared to the reference density, as discussed in Fig~\ref{rho}. The \redc{numerical results in GC-AdResS and the full-atom simulation agree now within $3\%$, which is highly satisfactory}.}
\label{tablecp2}
\end{center}
\end{table}
\redc{An additional test was done in order to prove that GC-AdResS satisfies a thermodynamic condition of a Grand Canonical ensemble in the thermodynamic limit. In fact in the thermodynamic limit the isothermal compressibility, $\kappa_{T}$, in a Grand Canonical ensemble, can be related to the fluctuations encoded in the particle number distributions \cite{salacuse}:
\begin{equation}
\rho k_{B}T\kappa_{T}=\frac{\langle N^{2} \rangle-\langle N \rangle^{2}}{\langle N\rangle}
\label{kt}
\end{equation}
where $\rho$ is the density of particles, $k_{B}$ the Boltzmann constant and, $T=298 K$, the temperature. The test was done for a system of 20000 molecules with a reservoir of 800000 (total number of molecules 100000) at a pressure of $1 atm$; in this case we obtained $\kappa_{T}=45.9\pm 1.2~~10^{6}(bar^{-1})$ which should be compared with the value of $44.6\pm 1.6~~10^{6}(bar^{-1})$ of the corresponding \redc{fully} atomistic system and with the value of about $45.25~~10^{6}(bar^{-1})$  \cite{exp1,exp2} of experiments and $44.0~~10^{6}(bar^{-1})$ from NPT simulations of SPC/E water \cite{exp2}; the overall accuracy is within $5\%$ (in the worst case), which can be considered a satisfactory result. It must also be underlined that an effective compressibility,  \redc{Eq.\ref{kt}, was found} to be the same in GC-AdResS and in the \redc{fully} atomistic simulation (see also Ref.\cite{prl12}).} Given the satisfactory tests for static properties, which prove that indeed the reservoir based on the local thermostat of GC-AdResS produces a Grand Canonical statistics,
we can now \redc{proceed with} the calculations of equilibrium time correlation functions where the notion of BL Liouville operator in the limit of a Grand Canonical Ensemble, comes into \redc{play} in order to provide theoretical solidity to the numerical calculations.
\subsection{Dynamic Properties}
\begin{figure}
   \centering
   \includegraphics[width=0.5\textwidth]{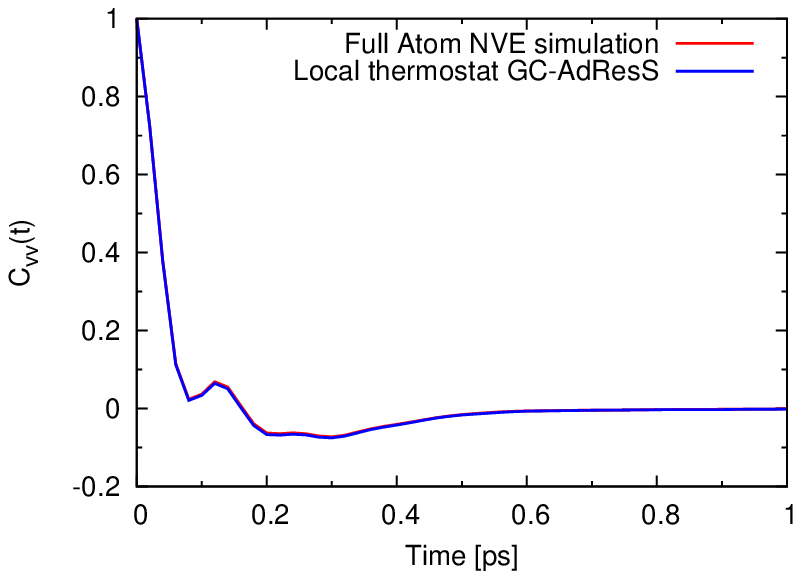}
   \includegraphics[width=0.5\textwidth]{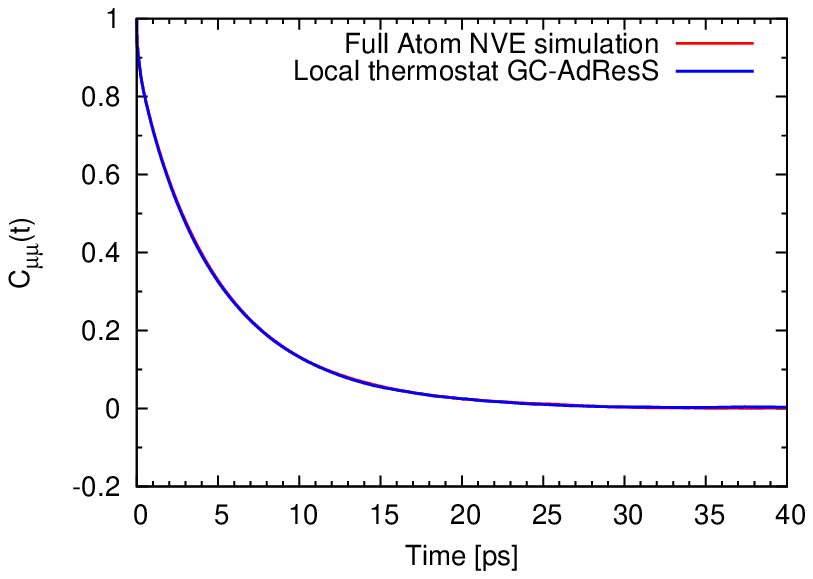}
   \includegraphics[width=0.5\textwidth]{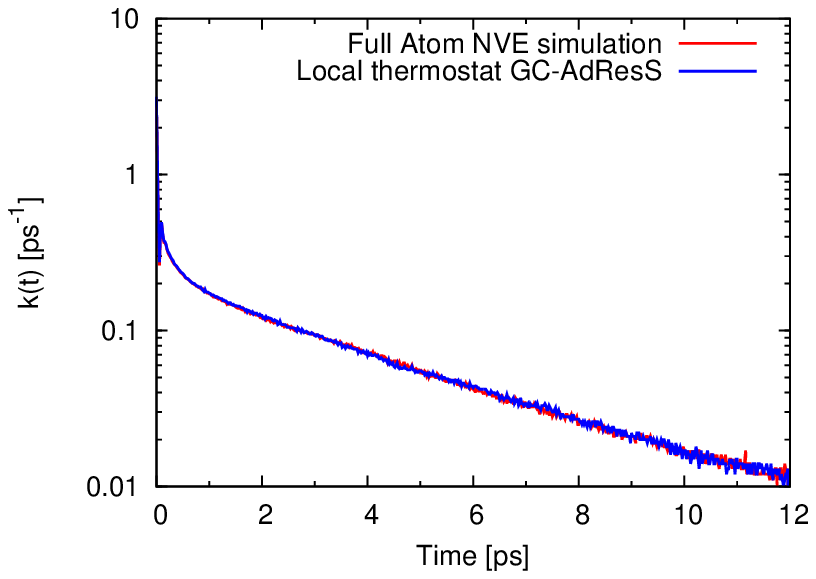}
   \caption{ Three relevant equilibrium time correlation functions for SPC/E water at room conditions calculated with GC-AdResS and  for an equivalent subsystem in a \redc{fully} atomistic NVE simulation.; as before, velocity-velocity autocorrelation function, $C_{VV}(t)$, (molecular) dipole-dipole autocorrelation function, $C_{\mu\mu}(t)$, reactive flux correlation function, $k(t)$ (semilogarithmic plot). The agreement between  GC-AdResS and the \redc{fully} atomistic simulation is highly satisfactory.}
   \label{etcorrfs}
 \end{figure}

\begin{figure}
   \centering
   \includegraphics[width=0.5\textwidth]{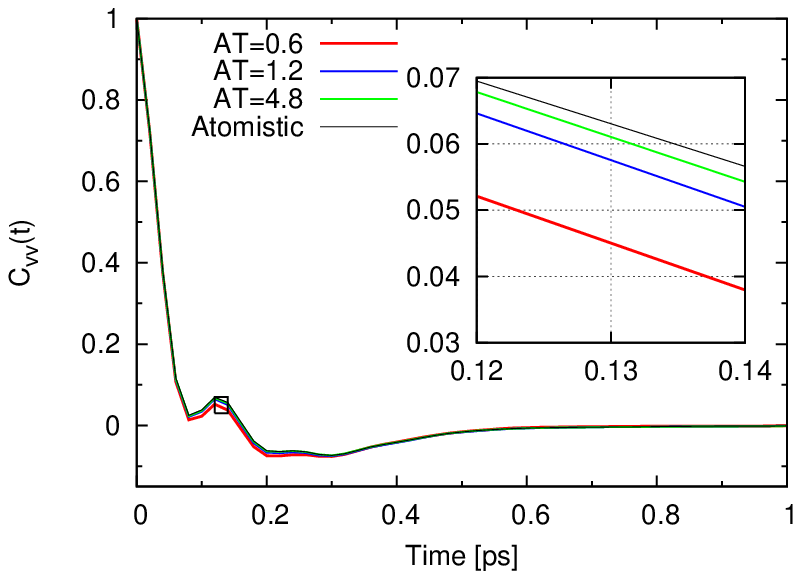}
   \includegraphics[width=0.5\textwidth]{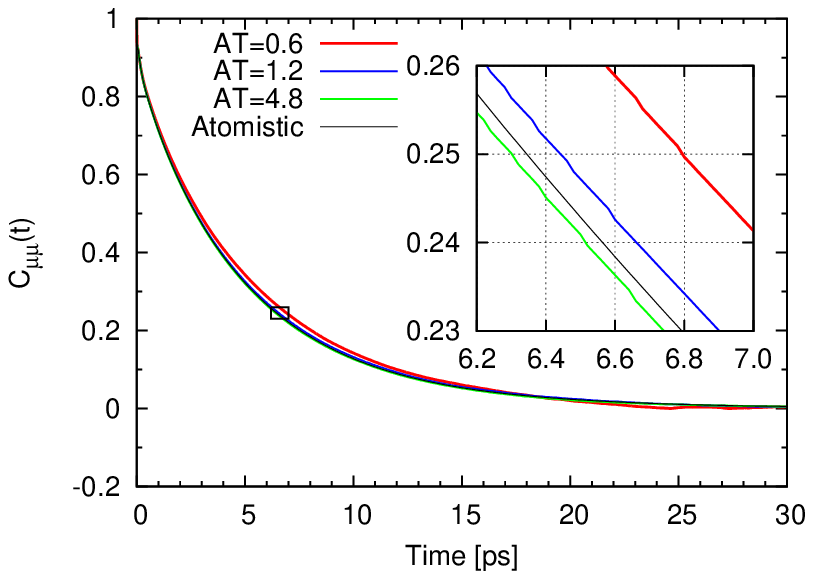}
   \includegraphics[width=0.5\textwidth]{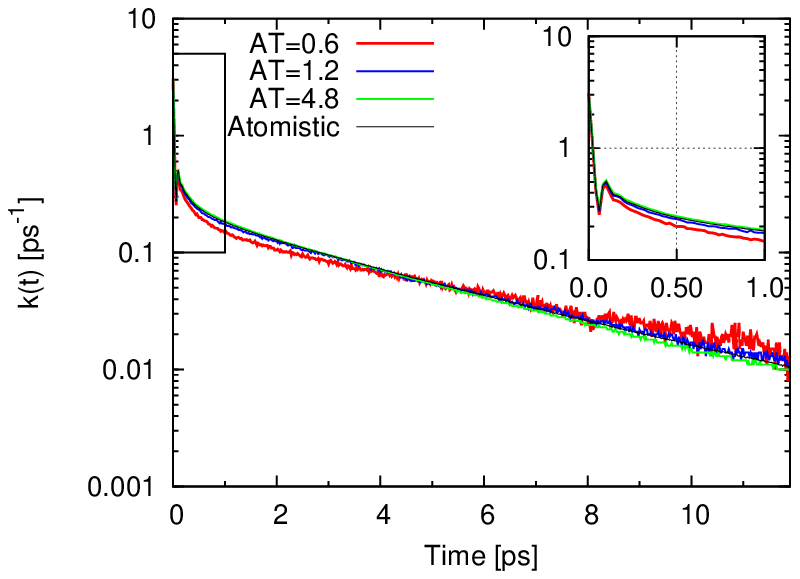}
   \caption{Systematic convergence of $C_{VV}(t)$, $C_{\mu\mu}(t)$ and $k(t)$ (semilogarithmic plot)  of GC-AdResS to the \redc{fully} atomistic NVE results calculated over the whole system.}
   \label{conv}
 \end{figure}

Here we report the numerical results of the application of GC-AdResS to the calculation of three relevant equilibrium time correlation functions for SPC/E water at room conditions. The GC-AdResS results are compared with the results obtained  for an equivalent subsystem in a \redc{fully} atomistic NVE simulation.
Fig.\ref{etcorrfs} shows the velocity-velocity autocorrelation function, $C_{VV}(t)$ (top), (molecular) dipole-dipole autocorrelation function, $C_{\mu\mu}(t)$ (middle), reactive flux correlation function, $k(t)$ (bottom); the agreement between GC-AdResS and the \redc{fully} atomistic NVE simulation is remarkable. This implies that the ``ideal'' reservoir of the GC-AdResS method is very close to \redc{the thermodynamic limit of a microscopic systems}.
A necessary condition of general validity of the concepts and calculations shown here is that as the AT region of GC-AdResS increases results must systematically converge to those obtained for the whole system of the \redc{fully} atomistic NVE simulation. This principle corresponds to the fact that in the thermodynamic limit all the ensembles are \redc{equivalent. Fig.\ref{conv}} shows the systematic convergence of the curves to the \redc{fully atomistic reference} as a function of the size of the AT subsystem of AdResS. A general remark valid when adaptive resolution is used as a multiscale technique rather than as Grand Canonical \redc{set-up} must be made: it must be noticed that the procedure defined above to calculate time correlation functions introduces a connection between the decay of a correlation function in time and the spatial locality of the process associated with such a decay. For example\redc{,} in dense gases decay times are relatively large, thus if the size of the atomistic region is too small, many molecules are likely to leave such region with the effect that the decay time would be shorter than the real one. In practical terms\redc{,} a way to probe whether or not our method captures a certain decay process is to perform a study where the size of the atomistic region is systematically varied and observe the convergence of the correlation function of interest. At the same time it must be also noticed that the connection between decay times and spatial locality is  not necessarily a limitation of the procedure, but actually represents one of its main conceptual advantages; in fact it allows to identify the essential (atomistic) degrees of freedom (in space and time) required for a certain process.

\section{Conclusions}
We have discussed the BL model as a \redc{prototypical theoretical construction} for describing the statistical mechanics of open systems.
Despite its conceptual solidity, the model has been not \redc{employed or discussed in connection with} the development of MD techniques with varying number of molecules.
\redc{As we have argued, however, the model turns out to be very useful as far as the conceptual validation of MD techniques is concerned.} We have discussed its connection to the GC-AdResS MD technique and used its principles to define equilibrium time correlation functions for system with a varying number of molecules. Numerical results for a relevant \redc{system, liquid} water at room conditions, are highly \redc{promising}. We have then discussed the computational efficiency and convenience of GC-AdResS. Given the technical robustness of GC-AdResS and its conceptual validation within the BL model, one can think, in perspective, to move forward and approach also systems \redc{out of} equilibrium, e.g. \redc{subject to} an external perturbation. For example, biomolecules in solution whose conformational dynamics is driven by an external (electric) field as in Ref.\cite{jctchan2}. The response of the system to an external perturbation requires a numerical technique similar to that employed in the calculation of equilibrium time correlation functions, \redc{moreover the region of microscopic interest is limited to the  first two-three solvation shells of the molecule, which is an ideal test case for a AdResS-like technique.} The study of open systems is gaining popularity and the \redc{development of} techniques which are both computationally efficient and theoretical well founded is a necessity of modern research in the field of molecular simulation; GC-AdResS is such an example.

\section*{Acknowledgments}
We thank Giovanni Ciccotti for uncountable clarifying discussions about the problem of Liouville Theorem in open systems and Luca Ghiringhelli and Matej Praprotnik for a critical reading of the manuscript.  This work was supported by the Deutsche Forschungsgemeinschaft (DFG) partially with the Heisenberg grant (grant code DE 1140/5-2) provided to L.D.S and partially with the grant CRC 1114 provided to L.D.S and C.H.. The  DFG grant (grant code DE 1140/7-1) associated to the Heisenberg grant for A.G. is also acknowledged. H.W. Acknowledges support from the National High Technology Research and Development Program of China under Grant No. 2015AA011201. Calculations were performed using the computational resources of the North-German Supercomputing Alliance (HLRN), project {\bf bec00100}.
\appendix
\section{Technical details}
\label{app:i}
All simulations are performed by home-modified GROMACS \cite{gromacs}, and the thermodynamic force in AdResS simulations is obtained using VOTCA \cite{votca} package.
The SPC/E \cite{spce} water model used in all the simulations. The system contains 5000 water molecules and the dimensions of the system are $14.6 \times 3.2 \times 3.2$ $nm^{3}$. In AdResS simulations, 
the resolution of the molecules changes only in the x-direction as depicted in Fig.\ref{fig1}. Three different atomistic regions are used in AdResS simulations, whose sizes are $0.6 \times 3.2 \times 3.2$ $nm^{3}$, 
$1.2 \times 3.2 \times 3.2$ $nm^{3}$ and $4.8 \times 3.2 \times 3.2$ $nm^{3}$ (this latter being a worst-case scenario for the reservoir, still results are \redc{very promising}). The size of the hybrid region is kept same in all the three cases $2.9 \times 3.2 \times 3.2$ $nm^{3}$. The remaining system contains 
coarse-grained particles, which interact via generic WCA potential of the form:
\begin{equation}
U(r) = 4\epsilon\bigg[\bigg(\frac{\sigma}{r}\bigg)^{12} - \bigg(\frac{\sigma}{r}\bigg)^{6}\bigg]  + \epsilon,            r \leq 2^{1/6}\sigma
\end{equation}
The parameters $\sigma$ and $\epsilon$ in the current simulations are 0.30 nm and 0.65 kJ/mol respectively. The time step used in the simulations is 0.002 ps, and the coordinates and velocities are recorded 
after every 10 time steps, i.e. 0.02 ps. All simulations are performed at room temperature 298 K. The coarse-grained and the hybrid region in the AdResS system are coupled to a Langevin thermostat, whose time scale is 0.1 ps. The reaction field method \cite{reax1, reax2} is used 
for calculating the electrostatic interactions in the system, with dielectric constant $\epsilon_{RF} = \infty$, as this tends to give good energy conservation. The ``switch" cut-off method is used to treat 
the van der waals interactions. The cut-off radius for interactions is 1.2 nm. For a 1 ns full atomistic simulation (without any thermostat), the total energy obtained is $-195846$ kJ/mol and the drift is just 11.4 kJ/mol
,which is less 0.01\%. The dynamical results from this Micro-Canonical ensemble are compared with results from AdResS simulations. All the dynamical properties are computed from equilibrated trajectories of 1 ns in \redc{fully} atomistic and AdResS simulations.  The velocity 
autocorrelation function is defined as:
\begin{equation}
C_{VV}(t) = \frac{1}{N}\sum_{i=1}^{N}\frac{\langle v_{i}(t) \cdot v_{i}(0) \rangle}{\langle v_{i}(0) \cdot v_{i}(0) \rangle}
\end{equation}
where $\langle \cdot \rangle$ denotes the equilibrium average and $\langle v_{i}(t) \cdot v_{i}(0) \rangle$ computes 
the correlation between the velocities of $i^{th}$ molecule at time 0 and $t$. In this work, the velocity auto correlation function 
is calculated only for the oxygen atoms. In the same way, the dipole auto correlation function is defined as:
\begin{equation}
C_{\mu\mu}(t) = \frac{1}{N}\sum_{i=1}^{N}\frac{\langle \mu_{i}(t) \cdot \mu_{i}(0) \rangle}{\langle \mu_{i}(0) \cdot \mu_{i}(0) \rangle}
\end{equation}
where  $\langle \mu_{i}(t) \cdot \mu_{i}(0) \rangle$ computes the correlation between the dipole moment of $i^{th}$ molecule at time 0 and $t$. 
In the current implementation of AdResS, the electrostatic interactions are calculated by short ranged reaction field method. The 
dipole auto correlation function results are consistent with the \redc{fully} atomistic simulation, also using reaction-field. We also tested Particle Mesh Ewald (PME) \cite{pme} as an alternative approach 
to compute coulomb interactions and calculated the dipole auto correlation function in a \redc{fully} atomistic simulation, and found that the results were identical.  
The reactive flux hydrogen bond correlation\cite{chandler1, chandler2, stanley1, stanley2} function is defined as: 
\begin{equation}
k(t) = -dC_{HH}/dt
\end{equation}
where $C_{HH}(t)$ is the hydrogen bond autocorrelation function defined as:
\begin{equation}
C_{HH}(t) = \frac{\langle h(0) \cdot h(t) \rangle}{\langle h \rangle}
\end{equation}
Here $h$ is the hydrogen bond population operator for a particular pair of molecules. It is assigned a value 
'1', if there is a hydrogen bond between this pair, otherwise a value '0'.  
The criteria for considering a hydrogen bond between two water 
molecules is (1) inter oxygen distance is less than 0.35 nm and (2) the $O-H \ldots O$ angle is smaller than $30^{o}$. 
The function $C_{HH}(t)$ is the conditional probability that a hydrogen bond between a pair of molecules is present 
at time 't', given that it was present at time zero. 
In both the \redc{fully} atomistic and AdResS simulations, first $C_{HH}(t)$ was calculated and then $k(t)$ was obtained by 
taking numerical derivative of $C_{HH}(t)$, using a time step of 0.02 ps.

\appendix
\section{Thermodynamic Fluctuations}
\label{app:j}
The following thermodynamic quantities are analyzed in this work:
\begin{equation}
Var(E) = \frac{\langle E^{2} \rangle - \langle E \rangle^{2}}{\langle E \rangle}
\end{equation}
and 
\begin{equation}
CoVar(N,E) = \frac{\langle NE \rangle - \langle N \rangle  \langle E \rangle}{\langle N \rangle}
\end{equation}
where $Var(E)$ is the variance in the total energy of the molecules in the atomistic subregion
in AdResS and an equivalent subregion in the full-atom simulations, $CoVar(N, E)$ is the
covariance between the total energy of the molecules and number of molecules which are present
in the atomistic subregion in AdResS and an equivalent subregion in the full-atom simulations. The energy $E$ consists of the sum of the kinetic energy of the molecules in the region considered, plus the energy coming from the interactions of each molecule with all the other molecules of the region considered. The interactions with the reservoir, defined in the text, ``technical interactions'', are not counted, for consistency with the definition of reservoir in the BL model. The different properties are calculated from a 2 ns long trajectory. 
The error in the data was calculated using ``block-averaging" analysis.



\begin{thebibliography}{24}
\expandafter\ifx\csname natexlab\endcsname\relax\def\natexlab#1{#1}\fi
\expandafter\ifx\csname url\endcsname\relax
  \def\url#1{\texttt{#1}}\fi
\expandafter\ifx\csname urlprefix\endcsname\relax\def\urlprefix{URL }\fi
\bibitem{blatt}
J.M.Blatt, Progr.Theor.Phys. {\bf 22}, 745 (1959)
\bibitem{cell}
H.Quian, Annu.Rev.Phys.Chem. {\bf 58}, 113 (2007)
\bibitem{leb1}
J.L.Lebowitz and P.G.Bergmann, Phys.Rev. {\bf 99}, 578 (1955)
\bibitem{leb2}
J.L.Lebowitz and P.G.Bergmann, Annals of Physics, {\bf 1}, 1 (1957)
\bibitem{leb3}
J.L.Lebowitz and A.Shimony, Phys.Rev. {\bf 128}, 1945 (1962)
\bibitem{jmp}
G.G.Emch and G.Sewell, J.Math.Phys. {\bf 9}, 946 (1968)
\bibitem{peier}
W.Peier, Physica, {\bf 57}, 565 (1971)
\bibitem{pettit1}
C.J.Lynch and B.M.Pettitt, J.Chem.Phys. {\bf 107}, 8594 (1997)
\bibitem{entropy}
L.Delle Site, Entropy, {\bf 16}, 23 (2014)
\bibitem{annurev}
M.Praprotnik, L.Delle Site and K.Kremer, Annu.Rev.Phys.Chem. {\bf 59}, 545 (2008)
\bibitem{parrix}
C.Perego, M.Salvalaglio, M.Parrinello, {\it ``Molecular Dynamics Simulations of Solutions at Constant Chemical Potential''}, arXiv:1501.07825 (2015)
\bibitem{seke}
J.Seke, Phys.Rev.A, {\bf 21}, 2156 (1980)
\bibitem{tuck}
M.E.Tuckerman, ``{\it Statistical Mechanics: Theory and Molecular Simulation}'';
Oxford University Press, New York 2010
\bibitem{tech}
I.V.Ovchinnikov and D.Neuhauser, J.Chem.Phys. {\bf 122}, 024707 (2005)
\bibitem{widom}
B.J.Widom, J.Chem.Phys. {\bf 39}, 2808 (1963)
\bibitem{flor}
H.Eslami and F.M\"{u}ller-Plathe, J.Compt.Chem. {\bf 28}, 1763 (2007)
\bibitem{pettitt}
T.Cagin and B.M.Pettitt, Mol.Sim. {\bf 6}, 5 (1991); T.Cagin and B.M.Pettitt, Mol.Phys. {\bf 72}, 169 (1991); S.Weerasinghe and B.M.Pettitt, Mol.Phys. {\bf 82}, 897 (1994); J.Li, T.Cagin and B.M.Pettitt, J.Chem.Phys. {\bf 96}, 1333 (1992)   
\bibitem{lo-palm}
B.J.J.Palmer and C.Lo,  J.Chem.Phys. {\bf 101}, 10899 (1994); C.Lo and B.J.J.Palmer, J.Chem.Phys. {\bf 102}, 925 (1995)
\bibitem{difpet}
T.Kuznetsova and B.Kvamme, Mol.Phys. {\bf 97}, 423 (1999).
\bibitem{adress}
M.Praprotnik, L.Delle Site and K.Kremer, J.Chem.Phys. {\bf 123}, 224106 (2005)
\bibitem{jctchan}
H.Wang, C.Sch\"{u}tte and L.Delle Site, J.Chem.Th.Comp. {\bf 8}, 2878 (2012)
\bibitem{MeynTweedie}
P. Meyn and R.L. Tweedie, \emph{Markov chains and stochastic stability}, Springer, London (1993) 
\bibitem{TheilLeimkuhler2009}
B. Leimkuhler, E. Noorizadeh, and F. Theil, A Gentle Stochastic Thermostat for Molecular Dynamics, J. Stat. Phys. {\bf 135}, 261 (2009) 
\bibitem{prx}
H.Wang, C.Hartmann, C.Sch\"{u}tte and L.Delle Site, Phys.Rev.X {\bf 3}, 011018 (2013)
\bibitem{ensing}
B.Ensing, S.O. Nielsen, P.B. Moore, M.L. Klein, and M.Parrinello, J.Chem.Th.Comp. {\bf 3}, 1100 (2007)
\bibitem{kurt}
R.Potestio, S.Fritsch, P.Espanol, R.Delgado-Buscalioni, K.Kremer, R. Everaers and D.Donadio, Phys.Rev.Lett. {\bf 110}, 108301 (2013)
\bibitem{truh}
A.Heyden and D.G.Truhlar, J.Chem.Th.Comp. {\bf 4}, 217 (2008)
\bibitem{note1}
In order to convert specific adaptive methods to a Grand Canonical-like \redc{set-up} (in most of such techniques) one must go beyond the request of having a global Hamiltonian with a physical meaning. In fact in the Hamiltonian and Lagrangian based techniques, one can find statements justifying as {\it ``realistic physical process''} the artificial/technical process of changing resolution of a molecule. This in turn led the adaptive idea to be conceptually forced in a Canonical or Microcanonical ensemble only for the sake of familiarity with standard MD approaches. A global Hamiltonian does not make the conceptual derivation more rigorous than a forced-based approach, in fact it leads to implicit violations of basic principles of statistical mechanics and/or an increase of computational costs (i.e., lack of energy conservation \cite{ensing,prlcomm}, thermodynamic state-dependent Hamiltonians employed for first principles statistical mechanics analysis \cite{hadstat}, increase of number of interactions as a function of the size of the system which makes the number of calculations impossible even for futuristic computers \cite{truh}). It is our opinion that the essential point is that the change of resolution is not a realistic physical process thus there is not a physics of reference against which to compare the correctness of properties of molecules with hybrid resolution. The only known first principle Hamiltonians are those of the high resolution and of the coarse grained region; they have a particle-dependent form and are Hamiltonians typical of a Grand Canonical formulation. The hybrid Hamiltonian cannot be first principle because nature does not display changing resolution as a physical realistic process, thus a global Hamiltonian, while may be technically useful in some cases \cite{hadmc}, is conceptually artificial by definition and may be interpreted as a regression of the conceptual validity of the adaptive resolution method (if the global Hamiltonian is used for conceptual validation of the method). Instead, the perspective we propose in this paper is at the same time simple and unambiguous, that is, to treat the adaptive technique as a Grand Canonical-like \redc{set-up} and consider the interface region as an nonphysical filter with negligible (but numerically quantified) surface-like effects over the rest of the system.
\bibitem{jcpan}
A.Agarwal, H.Wang, C.Sch\"{u}tte and L.Delle Site, J.Chem.Phys. {\bf 141}, 034102 (2014)
\bibitem{prl12}
S.Fritsch, S.Poblete, C.Junghans, G.Ciccotti, L.Delle Site and K.Kremer, Phys.Rev.Lett. {\bf 108}, 170602 (2012)
\bibitem{jctchan2}
H.Wang, C.Sch\"{u}tte, G.Ciccotti and L.Delle Site, J.Chem.Th.Comp. {\bf 10} 1376 (2014)
\bibitem{salacuse}
J.J.Salacuse, Physica A {\bf 387}, 3073 (2008)
\bibitem{exp1}
G.S.Kell, J.Chem.Eng.Data, {\bf 15}, 119 (1970)
\bibitem{exp2}
C.Huang, K.T.Wikfeldt, T.Tokushima, D.Nordlund, Y.Harada, U.Bergmann, M.Niebuhr, T.M.Weiss, Y.Horikawa, M.Leetmaa, M.P.Ljungberg, O.Takahashi, A.Lenz, L.Ojam\"{a}e, A.P.Lyubartsev, S.Shin, L.G.M.Pettersson, and A.Nilsson, Proc.Natl.Acad.Sci. {\bf 106}, 15214 (2009)
\bibitem{prlcomm}
M.Praprotnik, S.Poblete, L.Delle Site and K.Kremer,Phys.Rev.Lett. {\bf 107}, 099801 (2011) 
\bibitem{hadstat}
P. Espanol, R. Delgado-Buscalioni, R. Everaers, R. Potestio, D. Donadio and K. Kremer, J.Chem.Phys. {\bf 142}, 064115 (2015) 
\bibitem{hadmc}
R.Potestio, P.Espanol, R.Delgado-Buscalioni, R.Everaers, K.Kremer, D.Donadio, Phys.Rev.Lett. {\bf 111},  060601 (2013)
\bibitem{gromacs}
S.Pronk, S.Pall, R.Schulz, P.Larsson, P.Bjelkmar, R.Apostolov, M.R.Shirts, J.C.Smith, P.M.Kasson, D.van der Spoel, B.Hess, E.Lindahl, Bioinformatics {\bf 29}, 845 (2013).
\bibitem{votca}
V. R\"{u}hle, C. Junghans, A. Lukyanov, K. Kremer, and D. Andrienko, J. Chem. Theory Comput. {\bf 5}, 3211 (2009).
\bibitem{spce}
H. J. C. Berendsen, J. R. Grigera, and T. P. Straatsma, J. Phys. Chem. {\bf 91}, 6269 (1987)
\bibitem{reax1}
L. Onsager, Journal of American Chemical Society {\bf 58}, 1486 (1936)
\bibitem{reax2}
I.G. Tironi, R. Sperb, P.E. Smith, and W.F. van Gunsteren, J. Chem. Phys. {\bf 102}, 5451 (1995)
\bibitem{pme}
T. Darden, D. York and L. Pedersen, J. Chem. Phys. {\bf 98}, 10089 (1993)
\bibitem{chandler1}
A. Luzar and D. Chandler, Phys. Rev. Lett. {\bf 76}, 928 (1995)
\bibitem{chandler2}
A. Luzar and D. Chandler, Nature {\bf 379}, 55 (1996)
\bibitem{stanley1}
F. Starr, J. Nielsen, and H. Stanley, Phys. Rev. Lett. {\bf 82}, 2294 (1999)
\bibitem{stanley2}
F. Starr, J. Nielsen, and H. Stanley, Phys. Rev. E {\bf 62} 579 (2000)
\end{thebibliography}
\end{document}